# LC Oscillator Driver for Safety Critical Applications

Pavel Horsky

AMI Semiconductor Czech, Videnska 125, 619 00 Brno, Czech Republic
E-mail: pavel_horsky@amis.com

## Abstract

*A CMOS harmonic signal LC oscillator driver for automotive applications working in a harsh environment with high safety critical requirements is described. The driver can be used with a wide range of external components parameters (LC resonance network of a sensor). Quality factor of the external LC network can vary two decades. Amplitude regulation of the driver is digitally controlled and the DAC is constructed as exponential with piece-wise-linear (PWL) approximation. Low current consumption for high quality resonance networks is achieved. Realized oscillator is robust, used in safety critical application and has low EMC emissions.*

## 1. Introduction

The fast development of electronics in the last years with more and more advanced and precise sensors requires also more precise electronics to drive the sensors and measure their outputs. This driver is designed for a special sensor with an excitation coil used in an automotive application. The coil has to be driven with a harmonic current. To minimize the current consumption of the coil driver, the coil is a part of a LC resonance circuit and the driver has only to deliver losses in the resonance circuit. This harmonic signal is coupled into receiving coils and the amplitudes of the received signals are compared and then used to determine position of the sensor.

The main requirements for this oscillator driver are low current consumption for high quality LC circuits, very wide range of quality of the external LC circuit, and stable amplitude regulation. This sensor is used in safety critical applications, it can be utilized in dual oscillator systems and missing ground or supply on one of the oscillators should not influence the functionality of the other system. Detection of low amplitude, of missing oscillations or failure of external components is also required.

## 2. Oscillation condition

Let's assume an off-chip RLC network, where all losses are represented by serial resistance $R_S$ and an on chip oscillator driver in classical configuration with two transconductance $G_m$ stages (Fig. 1). When we take for simplicity $C = C_{osc1} = C_{osc2}$, then the oscillation condition (to keep stable oscillations) is

$$G_{m0} = R_S \cdot \frac{C}{L_{OSC}} = 2 \cdot \frac{R_S}{\omega^2 \cdot L_{OSC}^2} = \frac{1}{2} \cdot R_S \cdot \omega^2 \cdot C^2 \qquad (1)$$

To regulate the oscillation amplitude, non-linearity has to be inserted into the circuit. It can be created by limiting the output current of the drivers. The easiest approximation of the driver output current is shown in Fig. 2.

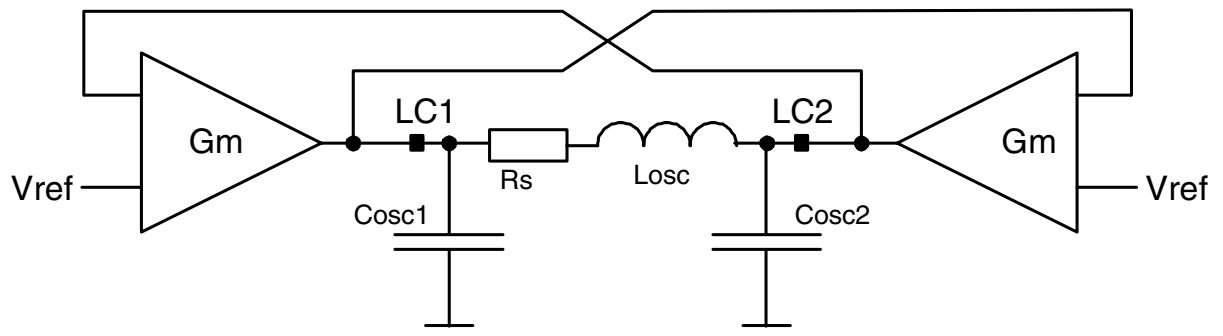

**Fig. 1 Oscillator driver**

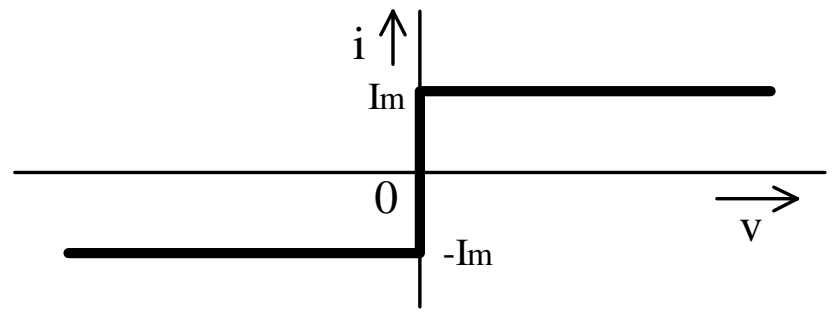

**Fig. 2 Driver current (static)**

The power dissipated in the RLC circuit is [1], [2]

$$P_{RLC} = V \cdot I = Rs \cdot I^2 = \frac{Rs}{2} \cdot \frac{C}{Losc} \cdot V^2 = \frac{1}{2} G_{m0} \cdot V^2 \qquad (2)$$

and the power delivered by the driver is

$$P_{drv} = V \cdot I_{drveq} = k \cdot V \cdot I_M \qquad (3)$$

where

$V$ is effective value of voltage between LC1 and LC2,

$I_{drveq}$ is equivalent driver output current $I_{drveq} = k \cdot I_M$

and $k$ depends on the VI characteristic of the driver, for linear approximation (Fig. 2) $k \approx 0.9$

Voltage of steady state oscillations is given by the formula

$$V = 2k \cdot \frac{L_{OSC}}{C \cdot R_S} \cdot I_M = 2k \frac{I_M}{G_{m0}} \qquad (4)$$

## 3. Amplitude regulation

Amplitude-voltage steps can be controlled by changing the current limitation (maximum driver current) by $\Delta I_M$. A linear voltage step requires an exponential current control.

$$\Delta V = 2k \cdot \frac{L_{OSC}}{C \cdot R_S} \cdot \Delta I_M = V \cdot \frac{\Delta I_M}{I_M} = V \cdot \delta I \qquad (5)$$

Since the Quality factor of the external RLC network can vary over two decades and low current consumption of the oscillator driver is required mainly for high quality resonance networks, a digital amplitude regulation with discrete current limitation steps was chosen. The current limitation for code $n$ is ($I_0$ is initial current step)

$$I_n = I_{n-1} \cdot (1 + \delta V) = I_0 \cdot (1 + \delta V)^n = I_0 \cdot M_n \qquad (6)$$

The required exponential function is approximated by a PWL function [4]. In Fig. 3 a 7-bit PWL approximated exponential DAC (corresponding to a 11-bit linear DAC) is shown.

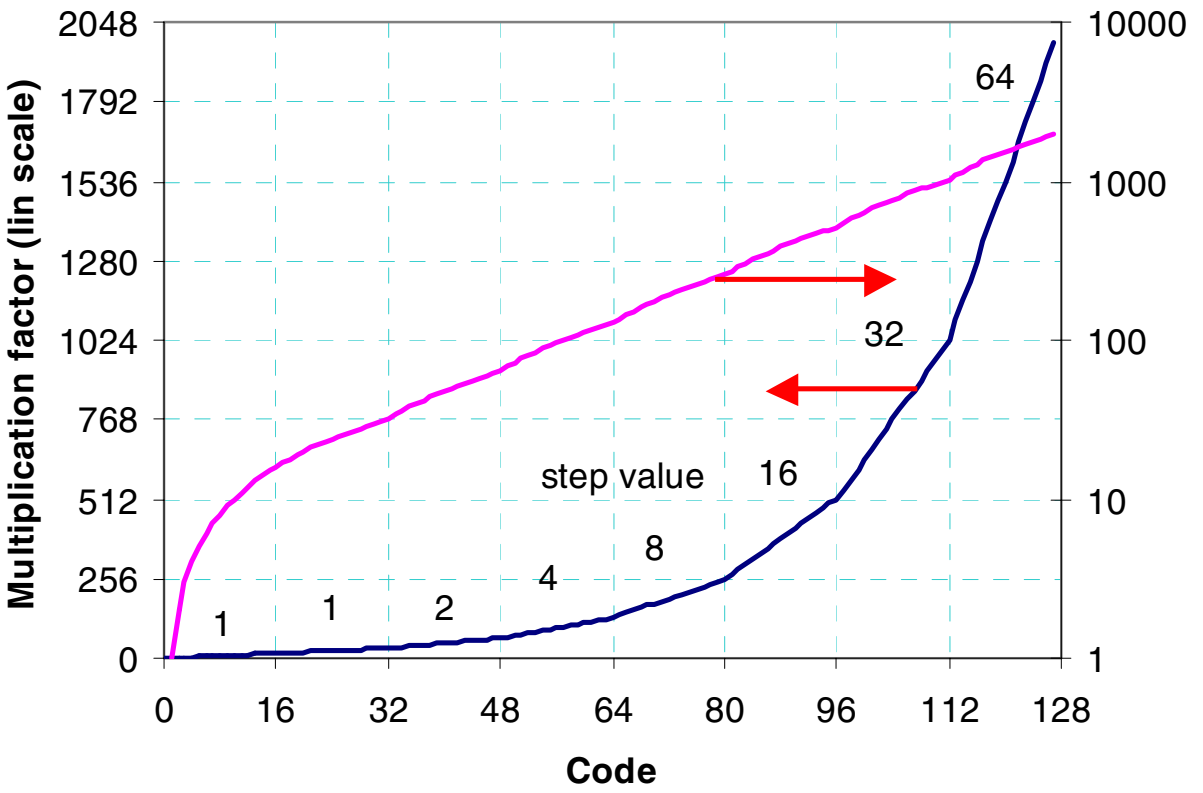

**Fig. 3 Current multiplication factor $M_n$ for a 7-bit PWL approximated exponential DAC**

The PWL approximation of the current control DAC, shown in Fig. 3, generates a variable amplitude step (Fig. 4). For codes above 16 the amplitude step varies between 3.23% and 6.25%. Losses in the resonance network and the driver regulation ensure that the amplitude regulation code remains above code 16.

## 4. Digital amplitude regulation loop

The oscillation amplitude is regulated by a state machine controlling the current limitation of the oscillator driver. The oscillator signal is full wave rectified, low-pass filtered and compared with two reference voltages by a window comparator. Window comparator was chosen to minimize the number of changes of the current limitation when working in steady state. Every 1 ms the oscillator driver current limitation is increased by one, decreased by one, or remains unchanged depending on the state of the window comparator.

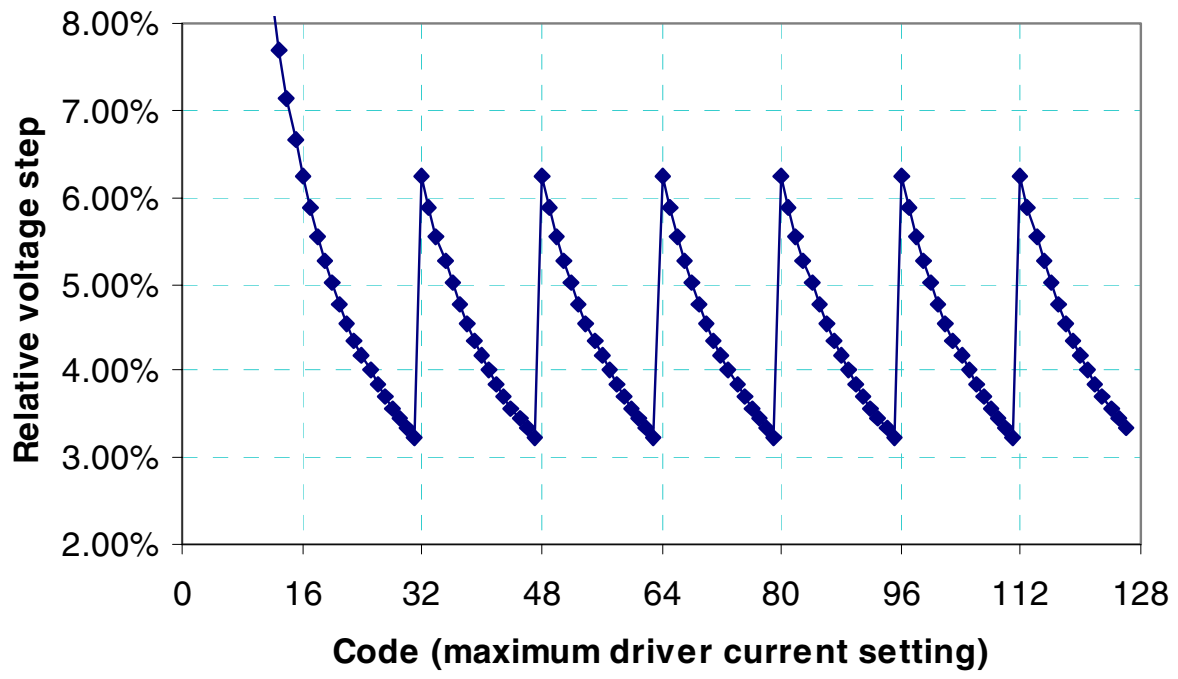

**Fig. 4 Relative voltage step as function of the current limitation code.**

The window for oscillator amplitude regulation is made wider than the maximum regulation step (6.25% for codes above 16). In this way, the regulation step can never jump over the window and cause regulation oscillations. The regulation loop allows a relaxed differential non-linearity of the DAC. The maximum step must only remain below a limit given by the regulation window and the converter can even be non-monotonic.

To reduce current consumption during start up (to approx. 40 % of the maximum current consumption), a power on reset signal sets the current limitation to code 105, which is lower than the maximum code, but is enough to start the oscillator even if maximum code for full amplitude is required. A few µs after startup an internal non-volatile memory is read and the code is set to a predefined value to speed up settling of the oscillator amplitude.

## 5. Realization

A main challenge for realization of the drivers is a wide dynamic range of output current (0:1984) and high speed (to limit losses the driver must be much faster than oscillation frequency, which is up to 5 MHz).

The full 7-bit scale of the DAC is divided into 8 ranges and in each of these ranges the output current step is constant (Fig. 3). Simplified block diagram of the oscillator driver is shown in Fig. 5. It consists of a prescale block (with scaling factors 1, 2, 4 and 8) delivering current *Iref2* into two complementary current mirrors. Both top and bottom current mirrors (Fig. 6) consist of two parts – one with fixed current outputs of 16, 16, 32 and 64 times *Iref2* and a second part with a 7-bit binary weighted current DAC with output current from 0 to 127 *Iref2*.

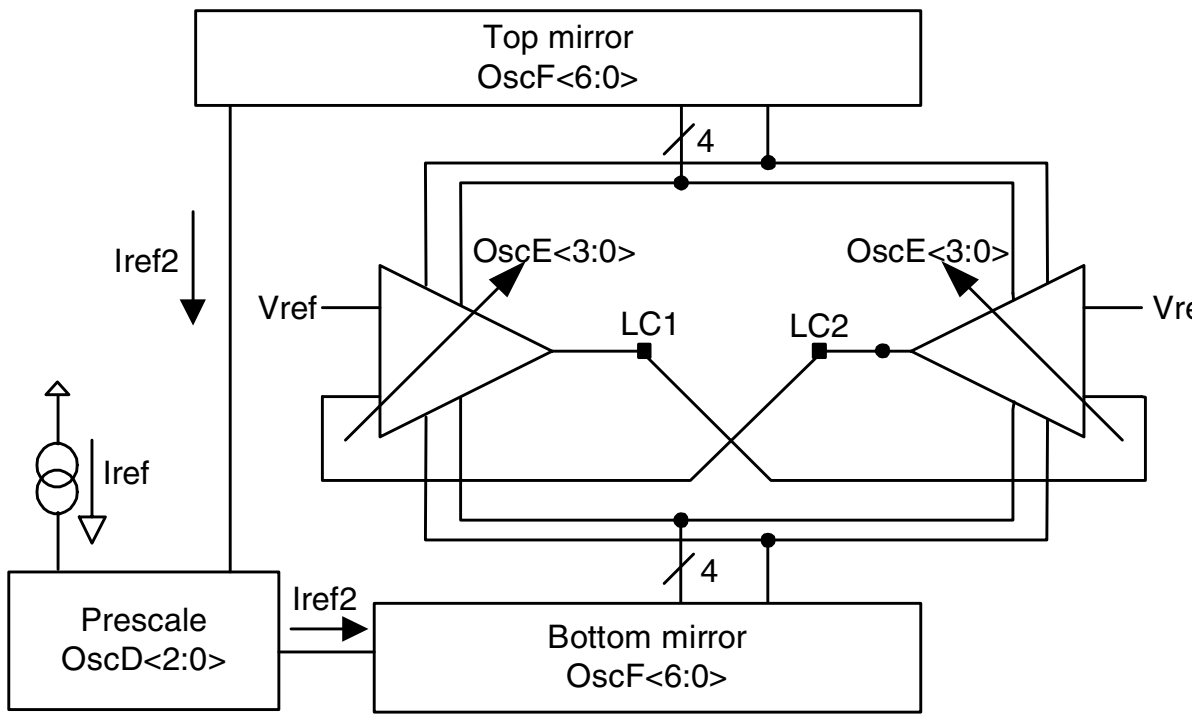

**Fig. 5 Simplified diagram of oscillator current limitation**

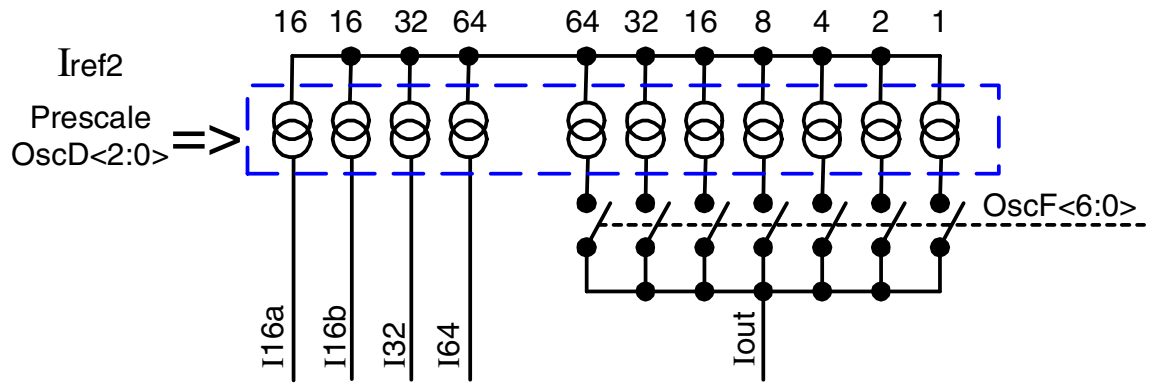

**Fig. 6 Block diagram of oscillator top current mirror**

Oscillator Gm block (Fig. 7) integrates two functions. When increasing the code, which increases current, it switches more fixed currents from top and bottom current mirrors to the output. The second function activates more output stages in parallel (see Table 1).

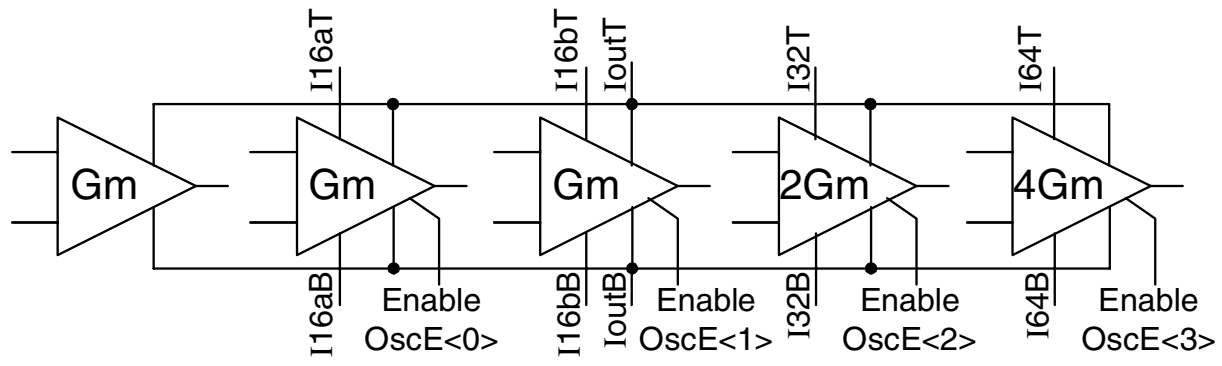

**Fig. 7 Simplified diagram of one oscillator Gm block (all inputs and outputs are connected in parallel)**

Setting of the current limitation is done using 3 independent control busses: prescaler bus OscD<2:0>, Gm-switching bus OscE<3:0> and current mirror bus OscF<6:0>. Table 1 shows how to generate the control signals. The output current can be calculated using following formula

$$I_{out} = I_{unit} \cdot (1 + OscD) \cdot \left[ OscF + 16 \cdot \left\{ OscE \bmod 2 + \mathrm{int}\frac{OscE}{2} \right\} \right]$$

## 6. Supporting circuits

The Vref point in Fig. 5 is connected to the middle of the supply voltage to control the DC operating point of the oscillator. To keep the DC operating point constant when the oscillator in dual system mode is overdriven from the other system, despite additional power consumption (typically 120 μA) a transimpedance amplifier is used with two output stages working in class A.

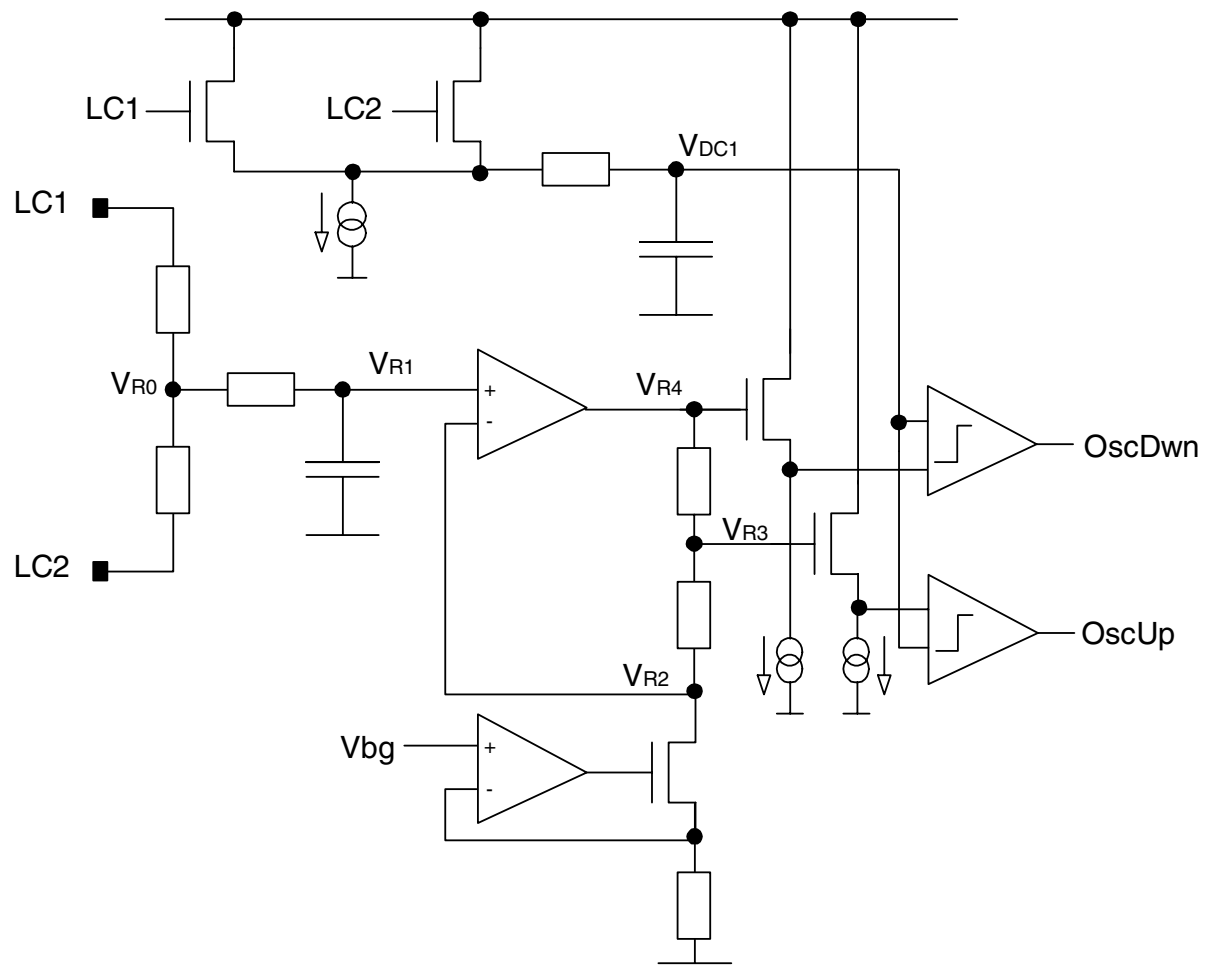

**Fig. 8 Amplitude detection circuits**

Amplitude detection is done by a full wave rectification and filtering of the voltages on LC1 and LC2 pins (Fig. 8). $V_{DC1}$ is then compared with two reference voltages $V_{R3}$ and $V_{R4}$ in a window comparator. These references are created by adding a fraction of the bandgap reference voltage $V_{BG}$ to the filtered mid-point between the LC1, LC2 pins: $V_{R1}$.

## 7. Safety critical requirements

This oscillator driver is used in a family of automotive products with high safety critical requirements. It can be utilized in dual systems with two oscillators with one or two excitation coils or dual systems with one oscillator only. Deep failure mode effect analysis (FMEA) on design and system levels of this oscillator driver are performed. For every external error condition the application must remain safe, it means the system has to detect the failure and set outputs according to it. In redundant systems with two drivers if one of the oscillators fails the other one should not be influenced – see paragraph 8. On chip level failure effects, their test coverage and their detect ability are evaluated for each block. ESD protections for every pin are also examined.

Detection of the following error conditions is performed:

- *Missing oscillations*
  Detects hard failures of the oscillator e.g. open connection to the coil, short to ground or supply. A fast comparator is connected between the pins LC1 and LC2 to create a clock signal. A missing clock is detected by a time-out circuit.

- *Low amplitude of oscillations*
  Detects pure quality of the LC network, e.g. a short in the coil or increased serial resistance of the coil. For low amplitude detection the same principle as mentioned in paragraph 6 is used.

- *Asymmetry of amplitude between the LC1 and LC2 pins*
  Detects failures on capacitors connected to the LC1 and LC2 pins. If one of the external capacitors $C_{osc1}$ or $C_{osc2}$ is missing or defective, the amplitude of the oscillations will be different between the LC1 and LC2 pins. The middle point voltage $V_{R0}$ between the LC1 and LC2 pins (Fig. 8) is no longer a DC voltage. This is detected by synchronous rectification, filtering and comparison with a reference voltage.

On complete system level other detections are also performed, e.g. monitoring supply voltage and detection of a short between the oscillator coil and receiving coils (monitoring if dc level on receiving coils can be easy changed) etc.

## 8. Overdriving output without supply

The oscillator can be used in redundant systems with two drivers and two mutually coupled excitation coils (the two systems are running at the same frequency) – see Fig. 9. If one of the systems looses supply or ground connection, it cannot load the other system, which must remain working.

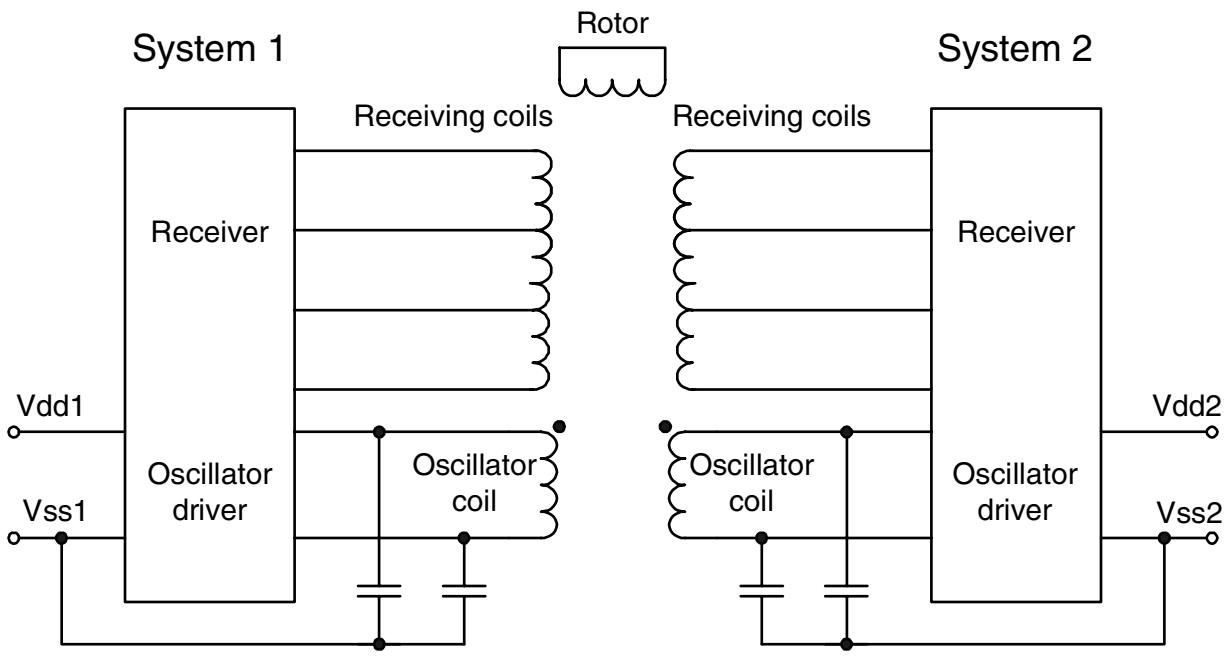

**Fig. 9 Redundant system**

The standard CMOS output driver (Fig. 10a) has intrinsically built in diodes, which will cause loading of the other oscillator if one system looses the supply voltage. In case an additional PMOS is used (Fig. 10b), the LCx pin can go negative, but the voltage range of the driver is limited (due to voltage needed to open MP1d).

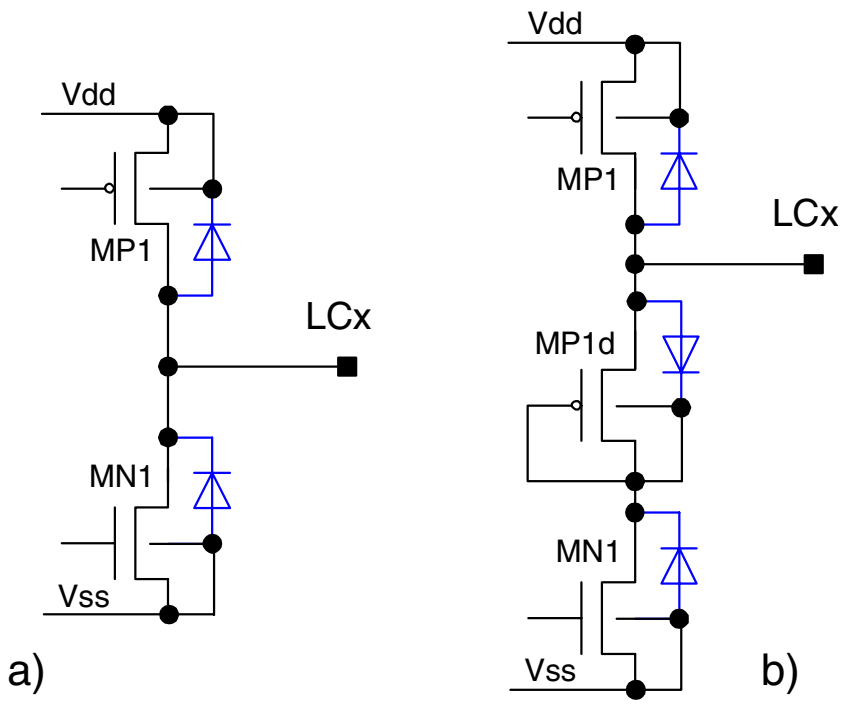

**Fig. 10 CMOS driver topology**

**Table 1 Coding of driver control signals**

| *DAC segment* | *7 bit input data B<6:0>* | | *Prescaler output Iref2* | *Active Gm stages* | **DAC output current** | | | **Control Signal Codes** | | |
|---|---|---|---|---|---|---|---|---|---|---|
| | | | | | *Step* | *Range min* | *Range max* | *Iref* | *Ibias OTA* | *DAC* |
| *-* | *MSBs* | *LSBs* | *units* | *-* | *units* | *units* | *units* | *OscD<2:0>* | *OscE<3:0>* | *OscF<6:0>* |
| 0 | 000 | B3 B2 B1 B0 | 1 | 1 | 1 | 0 | 15 | 000 | 0000 | 0 0 0 B3 B2 B1 B0 |
| 1 | 001 | B3 B2 B1 B0 | 1 | 2 | 1 | 16 | 31 | 000 | 0001 | 0 0 0 B3 B2 B1 B0 |
| 2 | 010 | B3 B2 B1 B0 | 2 | 2 | 2 | 32 | 62 | 001 | 0001 | 0 0 0 B3 B2 B1 B0 |
| 3 | 011 | B3 B2 B1 B0 | 2 | 3 | 4 | 64 | 124 | 001 | 0011 | 0 0 B3 B2 B1 B0 0 |
| 4 | 100 | B3 B2 B1 B0 | 4 | 3 | 8 | 128 | 248 | 011 | 0011 | 0 0 B3 B2 B1 B0 0 |
| 5 | 101 | B3 B2 B1 B0 | 4 | 5 | 16 | 256 | 496 | 011 | 0111 | 0 B3 B2 B1 B0 0 0 |
| 6 | 110 | B3 B2 B1 B0 | 8 | 5 | 32 | 512 | 992 | 111 | 0111 | 0 B3 B2 B1 B0 0 0 |
| 7 | 111 | B3 B2 B1 B0 | 8 | 9 | 64 | 1024 | 1984 | 111 | 1111 | B3 B2 B1 B0 0 0 0 |

To allow the LCx pin to go negative when the supply voltage is lost and to not change the voltage range of the driver we use the circuit in Fig. 11. When the chip is powered on, transistor MP6 is on and node Nbulk is shorted by MN6 to ground. To enable the driver, signals Ena and EnaN (inverted) are used.

Without supply, the voltage on Vdd is lower than 2 PMOS Vt needed to switch on MP6. MN6 is also off. For negative voltage on LCx pin (below NMOS Vt) transistors MN5 and MN3 are on connecting Nbulk and Ng1 to LCx potential. This ensures that MN1 is off. All PMOSes connected to LCx potential are also off (MP1, MP3, MP4) and there is no current flowing through the LCx pin. For positive overdrive on LCx bulk diode of MP1 is activated. MP3 is used to increase potential on node Ng2 to cancel the current path through MP1 (Itop is common for both LC1 and LC2 drivers). Special attention was paid to guarding of the driver, because it is used in a harsh environment.

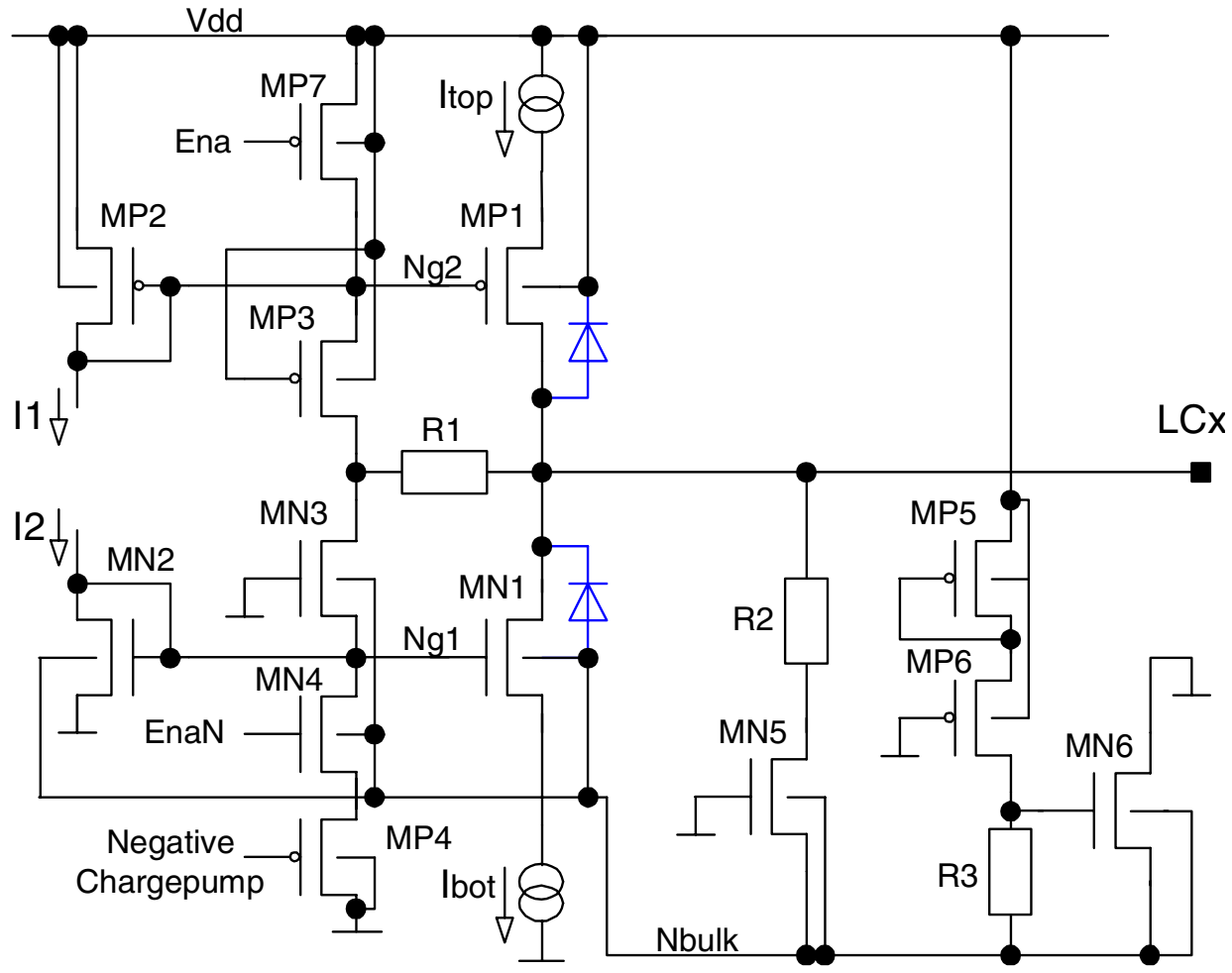

**Fig. 11 Output driver topology**

## 9. Realization and measurement results

The oscillator was processed in I3T80 technology - n-epi, 0.35 µm CMOS technology with single polysilicon and triple metal interconnect. Layout area of the driver is 0.22 mm$^2$ and area of the full oscillator including all detection blocks and 2 bond pads and ESD protections is 0.40 mm$^2$. The die photo with the different oscillator blocks is shown in Fig. 12.

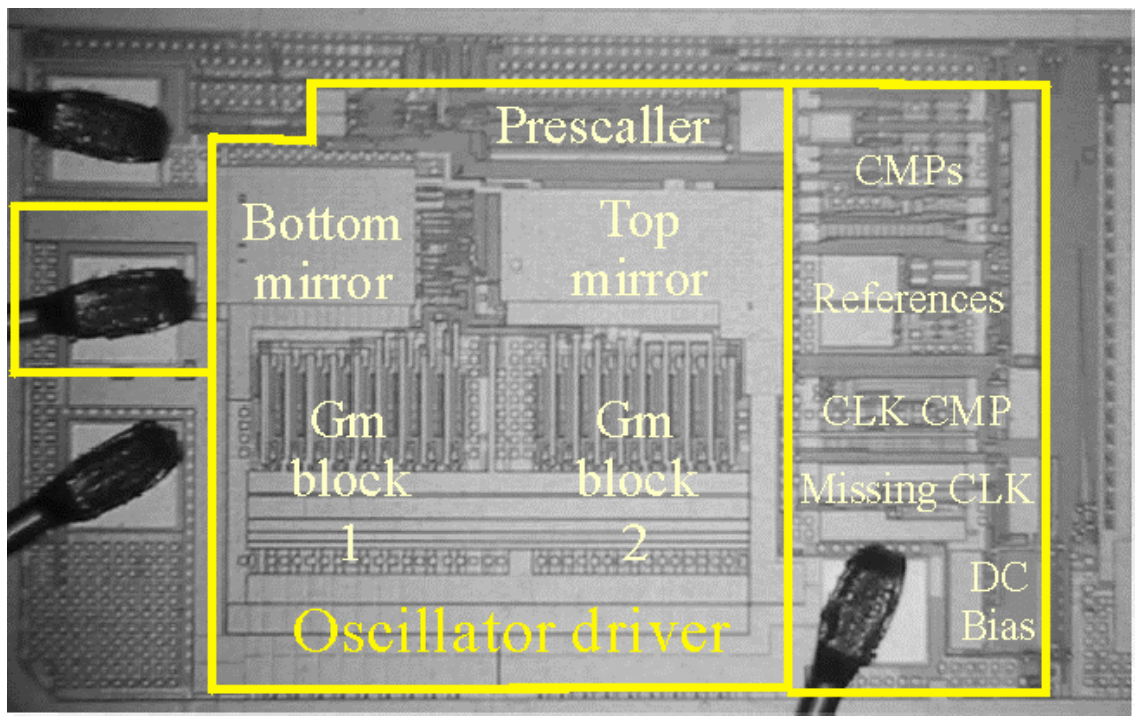

**Fig. 12 Photo of the realized oscillator driver**

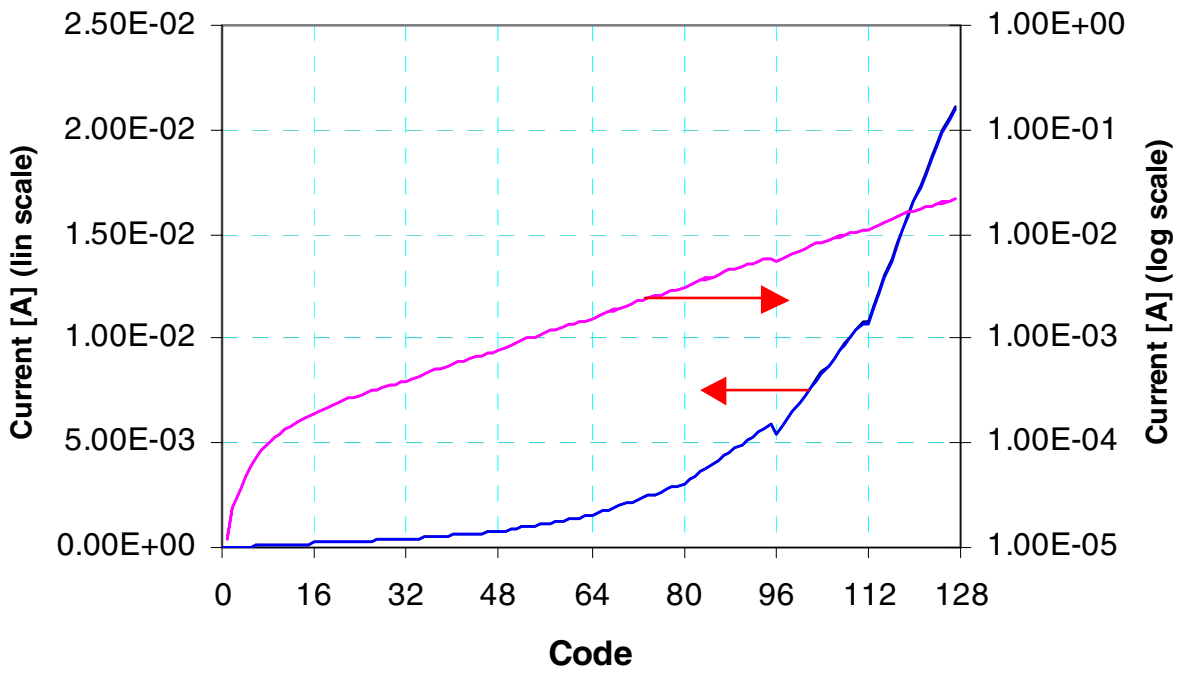

**Fig. 13 Measured current limitation**

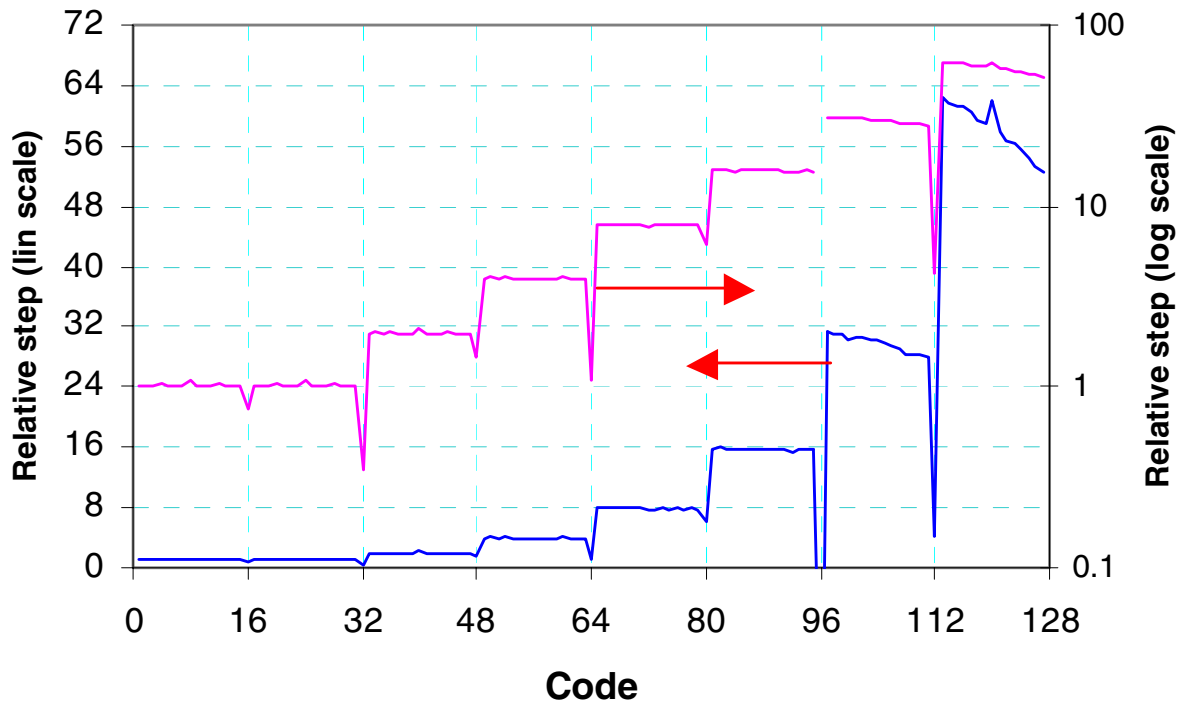

**Fig. 14 Measured relative current limitation step**

Fig. 13 shows measured current limitation of the driver (1 LSB is 12.5 µA). Measured relative current step is displayed in Fig. 14. Value for code 96 is negative (round 1 step in segment 7) and is removed for displaying in logarithmic scale. The DAC is non-monotonic at this code, but this is not a problem, because the regulation loop will regulate the amplitude as discussed before. Fig. 15 shows a detail of current regulation steps. Fast oscillator startup after enabling the driver is displayed in Fig. 16. Current consumption of the driver depends on the quality of the used LC resonance network and varies from 250 µA to 30 mA. The driver is designed for an oscillation frequency from 2 MHz to 5 MHz and works with wide range of external LC network parameters (equivalent trans-conductance up to around 10 mS is achieved for poor quality resonators and full amplitude).

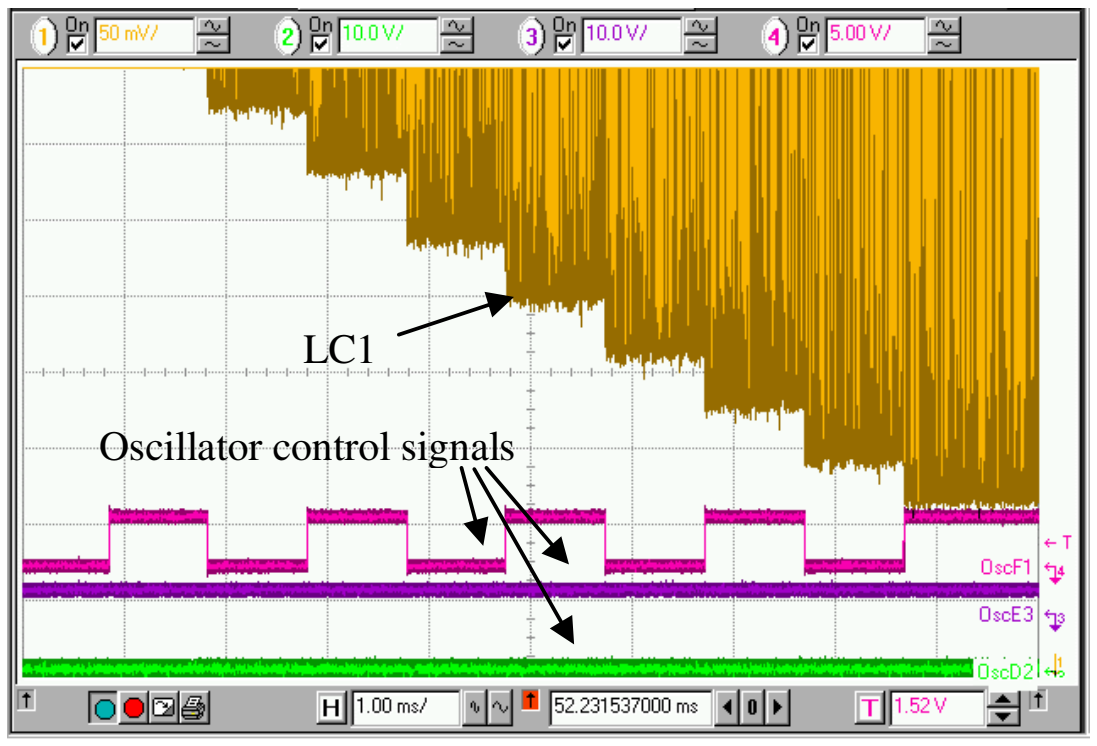

Fig. 15 Oscillator regulation steps (detail)

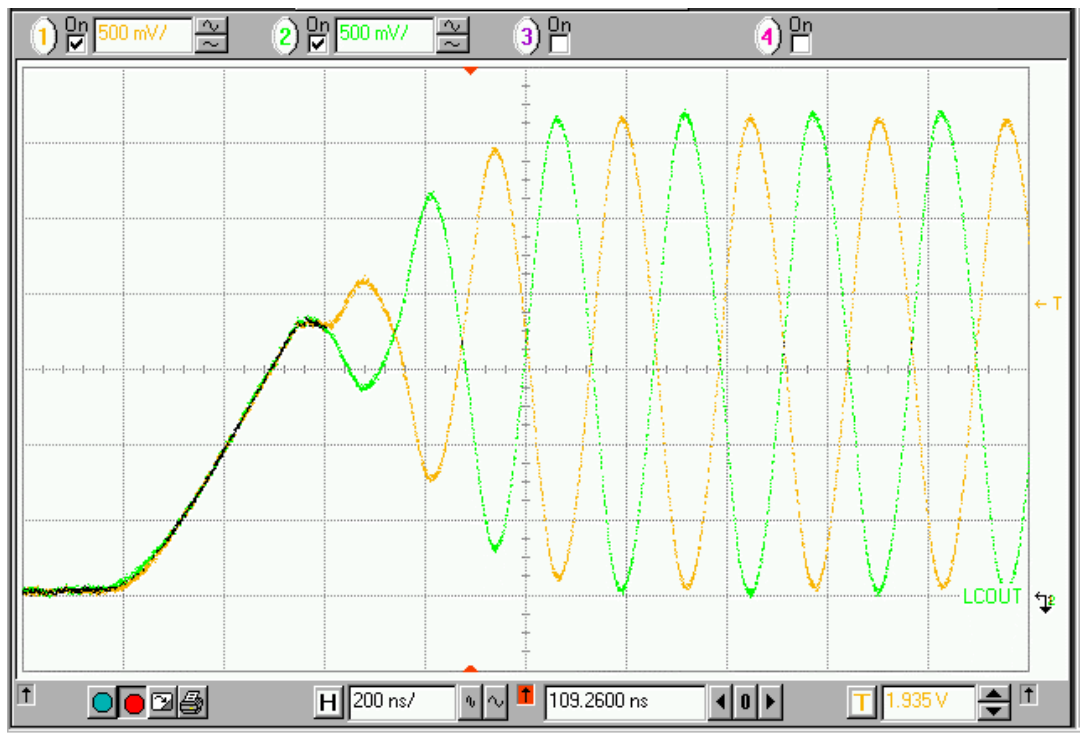

Fig. 16 Oscillator startup

All safety critical requirements described in paragraphs 7 and 8 where checked and the device is working properly. If low amplitude or missing oscillations are detected, the oscillator driver is set to maximum output current and outputs of the complete system are set to safe values.

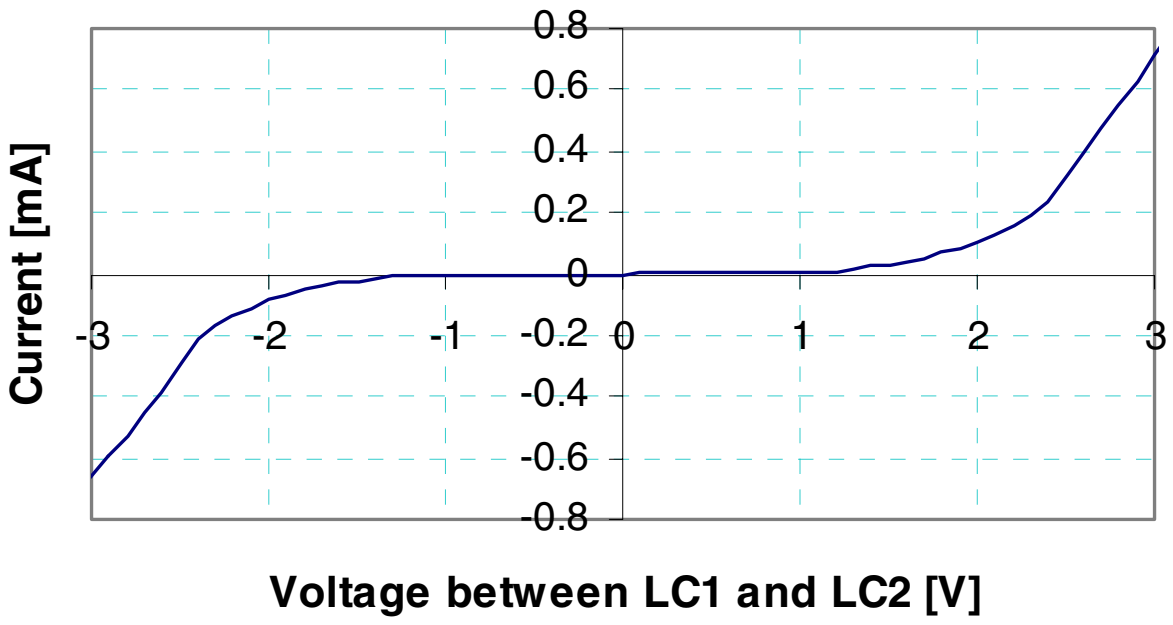

Fig. 17 Current through LC1,2 (Vdd is floating)

Fig. 17 and Fig. 18 show dc characteristic of the driver if the supply is missing. For maximum operating amplitude, which is 2.7 Vpp, the unsupplied system does not significantly influence the other system, which stays working.

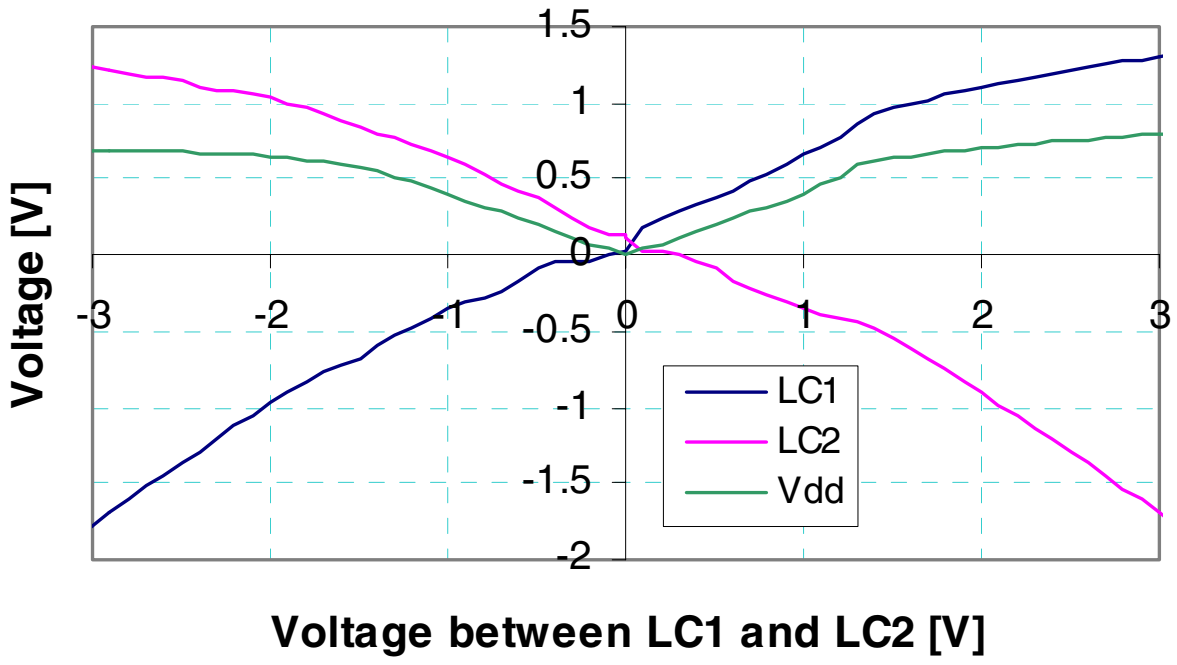

Fig. 18 Voltage on LC1,2 and Vdd (Vdd is floating)

## 10. Conclusions

A low current consumption LC oscillator driver for safety critical applications with a special type of exponential DAC for amplitude regulation was realized on small layout area in CMOS 0.35 µm technology. It is a robust driver capable of working with a wide range of external LC resonance network parameters. In case this oscillator is used in redundant systems with two separate supplies, the unique output driver topology ensures that the driver is not loading the other system if one supply voltage is missing. Deep analysis of failure mode effects was performed and the circuit is fulfilling the strong safety critical requirements.

### Acknowledgements

The author would like to thank H. Casier and J. Wright for their valuable comments to this paper.